\documentclass[prl,twocolumn,superscriptaddress,showpacs]{revtex4}
\usepackage{graphicx}
\begin{document}

\title{Prediction of the quantization of magnetic flux in double layer exciton superfluids}
\author{Louk Rademaker}
\email{rademaker@lorentz.leidenuniv.nl}
\affiliation{Institute-Lorentz for Theoretical Physics, Leiden University, PO Box 9506, Leiden, The Netherlands}
\author{Jan Zaanen}
\affiliation{Institute-Lorentz for Theoretical Physics, Leiden University, PO Box 9506, Leiden, The Netherlands}
\author{Hans Hilgenkamp}
\affiliation{Institute-Lorentz for Theoretical Physics, Leiden University, PO Box 9506, Leiden, The Netherlands}
\affiliation{Faculty of Science and Technology and MESA+ Institute for Nanotechnology, University of Twente, P.O. Box 217, 7500 AE Enschede, The Netherlands}
\date{\today}

\begin{abstract} 
Currently a way is lacking to detect unambiguously the possible phase coherence of an exciton condensate in an electron-hole double layer. Here we show that despite the fact that excitons are charge-neutral, the double layer exciton superfluid exhibits a diamagnetic response. In devices with specific circular geometry the magnetic flux threading between the layers must be quantized in units of $\frac{h}{e} \chi_m$ where $\chi_m$ is the diamagnetic susceptibility of the device. We discuss possible experimental realizations of the predicted unconventional flux quantization.
\end{abstract}


\maketitle

\begin{figure}
 \includegraphics[width=6.0cm]{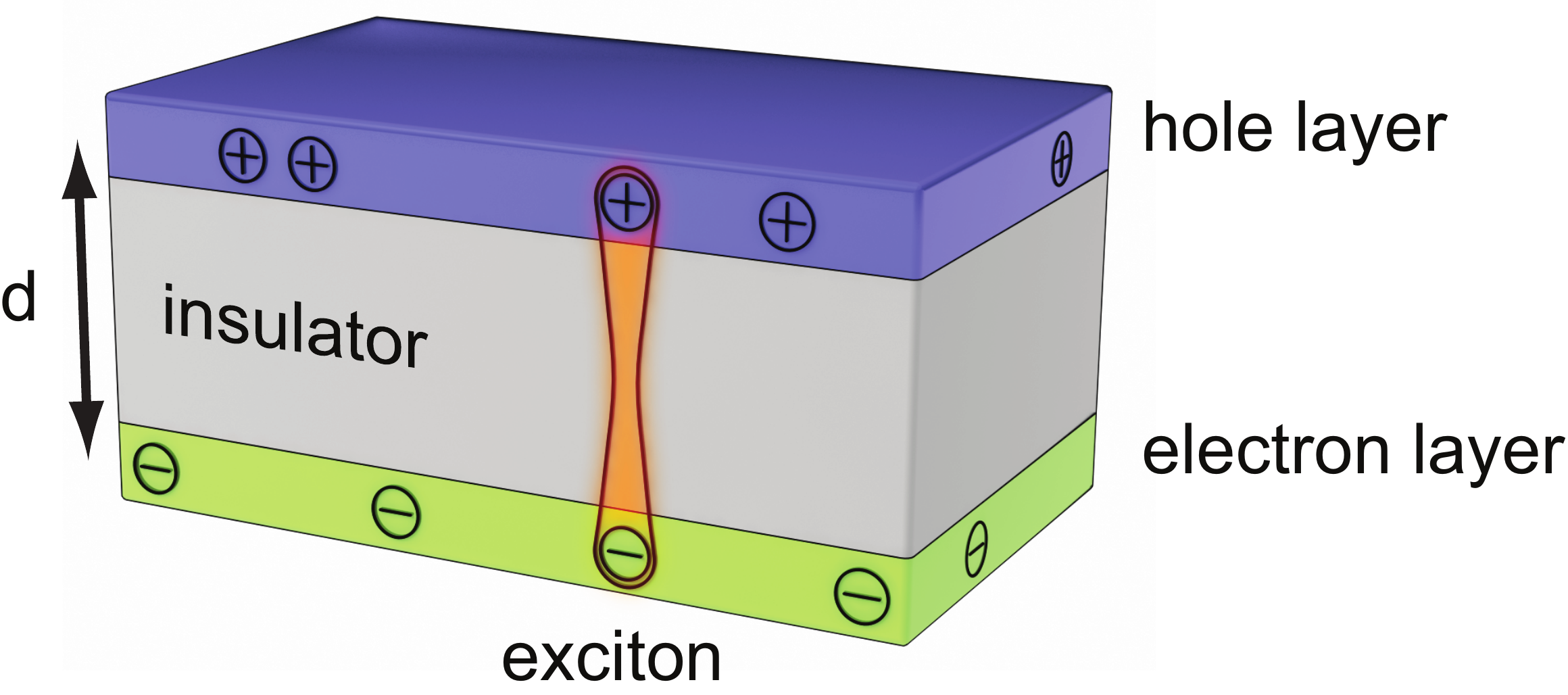}
 \caption{\label{fig1}Excitons in double layer devices. Double layer devices consist of an electron layer (in green) parallel to a hole layer (blue), separated by an insulating barrier (grey). Electron-hole attraction leads to the formation of excitons. At sufficiently low temperatures the excitons can form a Bose-Einstein condensate.}
\end{figure}

\emph{Introduction.}\textemdash It is well known that a quantization condition applies to the magnetic flux enclosed by a superconducting cylinder \cite{Deaver:1961p4961,Doll:1961p4835}. This effect is due to the coherence inherent to the superfluid phase, causing quantum-mechanical principles to become manifest in macroscopic objects \cite{JLeggett:2006p5020}. Such a superfluid phase is also predicted for excitons in spatially separated electron and hole layers \cite{Shevchenko:1976p4950,Lozovik:1976p4951,Moskalenko:2000p4767}. Excitons are bound pairs of electrons and holes and have a long history as optical excitations in semiconductors and insulators. Recent technological developments allowed realization of devices \cite{Butov:2003p4955,Eisenstein:2004p4770} that consist of a pair of two dimensional layers, where the electrons are confined to one layer and the holes to the other one as shown in Figure \ref{fig1}. When the interlayer distance is small enough (typically of order of 10 nanometer) the interlayer Coulomb interaction becomes strong enough to bind the electrons and the holes in interlayer excitons. An insulating barrier separates the layers to prevent annihilation of the excitons by tunneling. The layers themselves can be composed of semiconductor quantum wells \cite{Butov:2003p4955}, graphene sheets \cite{Lozovik:2008p4877,Zhang:2008p4895,Dillenschneider:2008p4896,Su:2008p4797,Kharitonov:2008p5044,Min:2008p4795}, complex oxides \cite{Pentcheva:2009p5025, Millis:2010p5052} or even topological insulators \cite{Seradjeh:2009p4980}.

\begin{figure}
 \includegraphics[width=8.6cm]{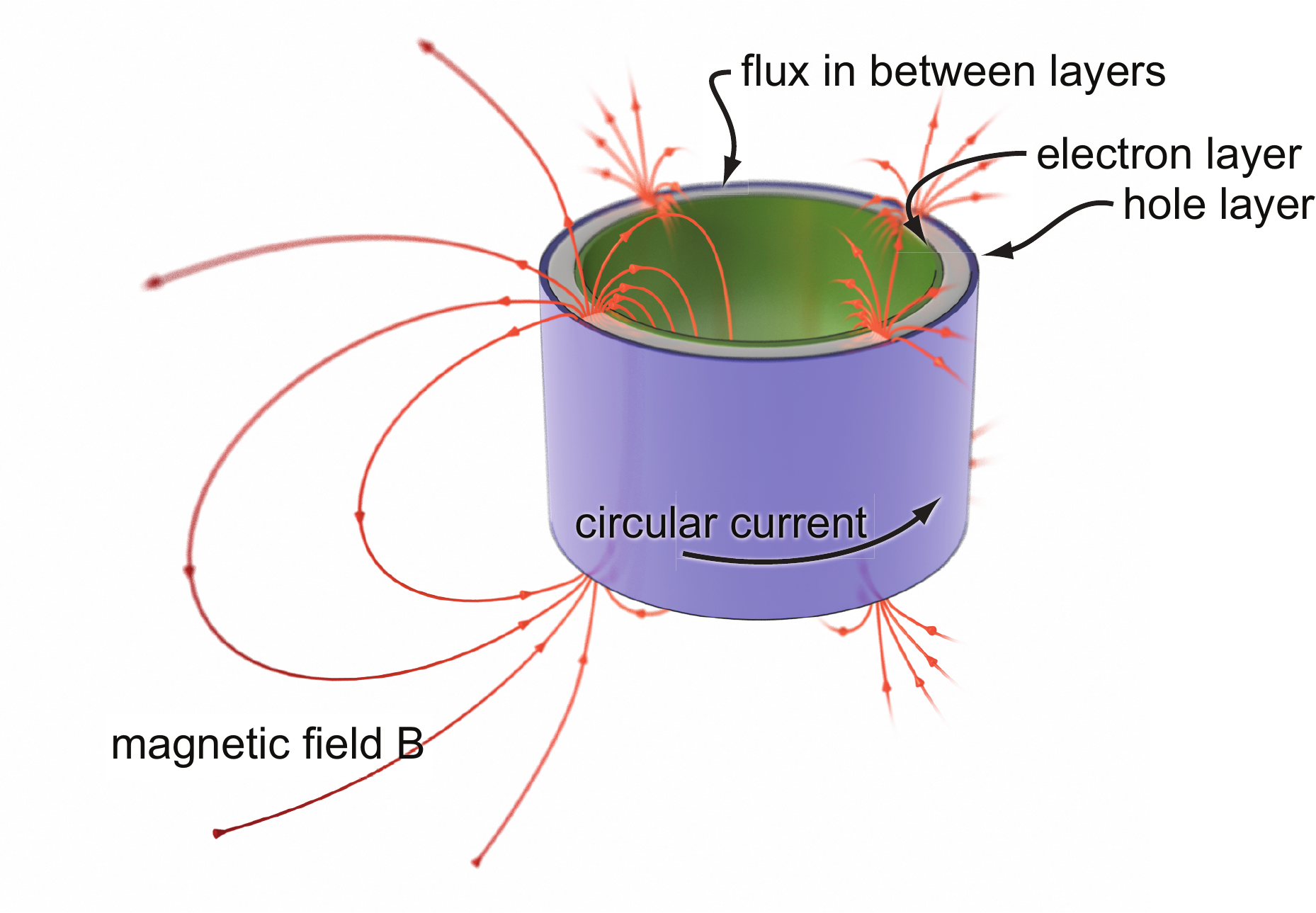}
 \caption{\label{fig2}Flux trapping in a cylindrical exciton superfluid. The proposed device consists of a concentric ring structure of radius $r$, composed of an electron layer (green) and hole layer (blue). Due to the macroscopic coherence of the exciton superfluid the angular current must be quantized. By application of an external axial magnetic field one can induce some number of current quanta. In absence of the external field, the current quanta remain which induces a magnetic field as shown (red lines). The trapped magnetic flux in between the layers must be quantized according to $\Phi = \frac{h}{e} \chi_m \; n$ where $\chi_m$ is defined in equation (\ref{ChiM}).}
\end{figure}

Excitons are bosons and at finite densities they should eventually form a Bose-Einstein condensate at sufficiently low temperatures. There are indications in several experiments \cite{Butov:2003p4955,Eisenstein:2004p4770} that exciton condensates were formed, but a way is lacking to detect unambiguously the onset of macroscopic superfluid coherence in these double layer exciton systems\cite{Snoke:2003p5046}. Here we predict an unconventional magnetic flux quantization effect to occur in double layer exciton superfluids, as shown in Figure \ref{fig2}, and we discuss designs for a device to measure this universal electromagnetic signature of the exciton Bose-Einstein condensate.

\emph{Ginzburg-Landau theory.}\textemdash Let us consider the Ginzburg-Landau order parameter theory for a double layer system. Since the direction of the electric dipole is fixed in the double layer geometry, the exciton superfluid is characterized by just a complex scalar order parameter field $\Psi (\vec{x}) \equiv |\Psi(\vec{x})| e^{i \phi(\vec{x})}$ along a 2D surface, the square of which gives the superfluid density $\rho(\vec{x}) = |\Psi (\vec{x})|^2$. For a charged superfluid/superconductor with boson charge $q$ electromagnetism is incorporated by replacing ordinary derivatives with covariant derivatives $\vec{D}$,
\begin{equation}
	\hbar \vec{D} = \hbar \vec{\nabla} + i q \vec{A} (\vec{x})
	\label{U1gauge}
\end{equation}
where $\vec{A} (\vec{x})$ is the vector potential. In the charge-neutral exciton superfluid the electron and hole constituents of an exciton form an electric dipole $e \vec{d}$ and consequently the covariant derivative associated with exciton matter must equal \cite{Balatsky:2004p3613}
\begin{equation}
	\hbar \vec{D} = \hbar \vec{\nabla} 
		+ i e \left[ \vec{A} (\vec{x} + \vec{d}/2) - \vec{A} (\vec{x}-\vec{d}/2) \right]
		\label{ExcitonCD}
\end{equation}
where the electron is positioned at $\vec{x}-\vec{d}/2$ and the hole at $\vec{x} + \vec{d}/2$. For small interlayer distance $\vec{d}$ the vector potential can be expanded in a Taylor series. In addition, since the vector potential $\vec{A}$ along the 2D superfluid surface is only sourced by in-plane currents, we can impose that the gradient of the vector potential component perpendicular to the surface is zero, 
\begin{equation}
	\left. \vec{\nabla}' 
		\left( \vec{d} \cdot \vec{A} (\vec{x}') \right) 
	\right|_{\vec{x'} = \vec{x}} = 0.
\end{equation}
This implies that the above vector potential difference can be written completely in terms of the 'real' magnetic field
\begin{eqnarray}
	&\vec{A} & (\vec{x} + \vec{d}/2) - \vec{A} (\vec{x}-\vec{d}/2)
	= \nonumber  \\ & &\left.
	- \vec{d} \times \sum_{k=0}^\infty \frac{1}{(2k+1)!} 
	\left( \frac{\vec{d}}{2} \cdot \vec{\nabla}' \right)^{2k} \vec{B} (\vec{x}') 
	\right|_{\vec{x}' = \vec{x}} .
\end{eqnarray}
Up to first order the exciton covariant derivative turns into
\begin{equation}
	\hbar \vec{D} = \hbar \vec{\nabla} 
		- i e \vec{d} \times \vec{B}.
	\label{SU2derivative}
\end{equation}
This is an interesting structure viewed from a theoretical perspective. Equation (\ref{SU2derivative}) corresponds to the covariant derivatives of a $SU(2)$ gauge theory with gauge fields $A_i^a = \epsilon^{iak} B_k$. Here the $SU(2)$ gauge fields are actually physical fields fixed by Maxwell's equations. Using the above considerations we can write down a general Ginzburg-Landau free energy
\begin{eqnarray}
	F [\Psi] &=& \int d^2x 
		\left[ \alpha |\Psi|^2 + \frac{1}{2} \beta |\Psi|^4 
			+ \frac{\hbar^2}{2m^*} (\nabla |\Psi|)^2 
		\right. \nonumber \\  &&	\left.
			+ \frac{1}{2m^*} \left[ \hbar \vec{\nabla} \phi 
			- e \vec{d} \times \vec{B} \right]^2 |\Psi|^2 
		+ d \frac{B^2}{2 \mu_0} \right].
	\label{FreeEnergy1}
\end{eqnarray}
The parameters $\alpha$ and $\beta$ can be written formally as a function of the superfluid density and the critical magnetic field $B_c$. Minimization of the free energy assuming a constant order parameter yields
\begin{eqnarray}
	\alpha &=& - d \frac{B_c^2}{\mu \rho}, \\
	\beta & = & - \frac{\alpha }{\rho}.
\end{eqnarray}

\emph{Electromagnetic response.}\textemdash The direct coupling to physical fields changes the rules drastically as compared to normal superconductors. We define the exciton supercurrent as the standard Noether current $\vec{j} \equiv \frac{\hbar \rho}{m^*} \vec{\nabla} \phi$ \cite{JLeggett:2006p5020}. Consequently, minimizing the free energy for a fixed applied magnetic field $\vec{B}$ perpendicular to the dipole moment yields the exciton supercurrent response
\begin{equation}
	\vec{j} 
	\equiv \frac{\hbar \rho}{m^*} \vec{\nabla} \phi 
	= \frac{\rho e}{m^*} \vec{d} \times \vec{B}.
	\label{Supercurrent}
\end{equation}
This result is closely related to spin superfluids \cite{Leurs:2008p4904} where a 'physical field' $SU(2)$ structure arises through spin-orbit coupling \cite{Goldhaber}. The analogue of Eq. (\ref{Supercurrent}) is the spin Hall equation \cite{Murakami:2003p4919} $j^i_j = \sigma_{\mathrm{s}} \varepsilon^{ijk} E_k \rightarrow \vec{j} = - \sigma \vec{d}_m \times \vec{E}$. We conclude that the spin-superfluid formed from magnetic dipoles is the electromagnetic dual of the exciton (electric dipole) superfluid. 

In the double layer system the electric charges forming the exciton dipoles are confined in the separate layers. Hence the exciton supercurrent can be decomposed into the separated electron and hole surface currents. According to Amp\`{e}re's law, a surface current induces a discontinuity in the magnetic field components parallel to the surface,
\begin{equation}
	\Delta \vec{B} ( \vec{x} ) = \mu_0 \vec{K} ( \vec{x} ) 
		\times \hat{n},
	\label{Ampere}
\end{equation}
where $\hat{n}$ is the normal vector to the surface and $\vec{K} (\vec{x})$ is an electric surface current density. Consequently, an exciton supercurrent reduces the magnetic field in between the electron and hole layer. The double layer therefore acts as a (non-perfect) diamagnet with magnetic susceptibility
\begin{equation}
	\chi_m = - \frac{e^2 \rho d \mu_0}{m^*}.
	\label{ChiM}
\end{equation}
For typical parameters $\rho = 0.4$ nm$^{-2}$, $d = 20$ nm and $m^* = 2 m_e$, the magnetic susceptibility equals $\chi_m = - 10^{-4}$,
comparable to what is found in diamagnets like gold or diamond.

\begin{figure}
 \includegraphics[width=5.00 cm]{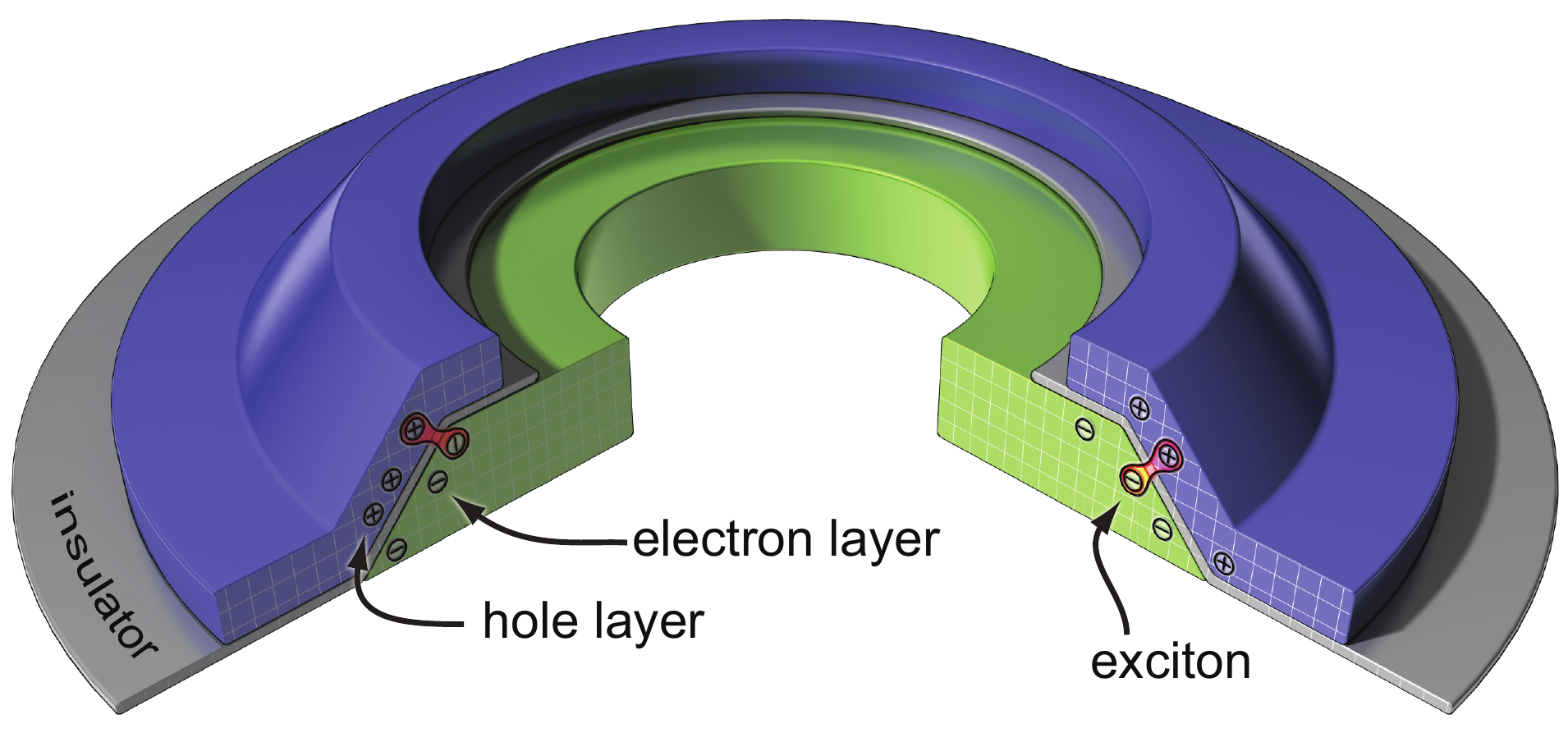}
 \includegraphics[width=3.50 cm]{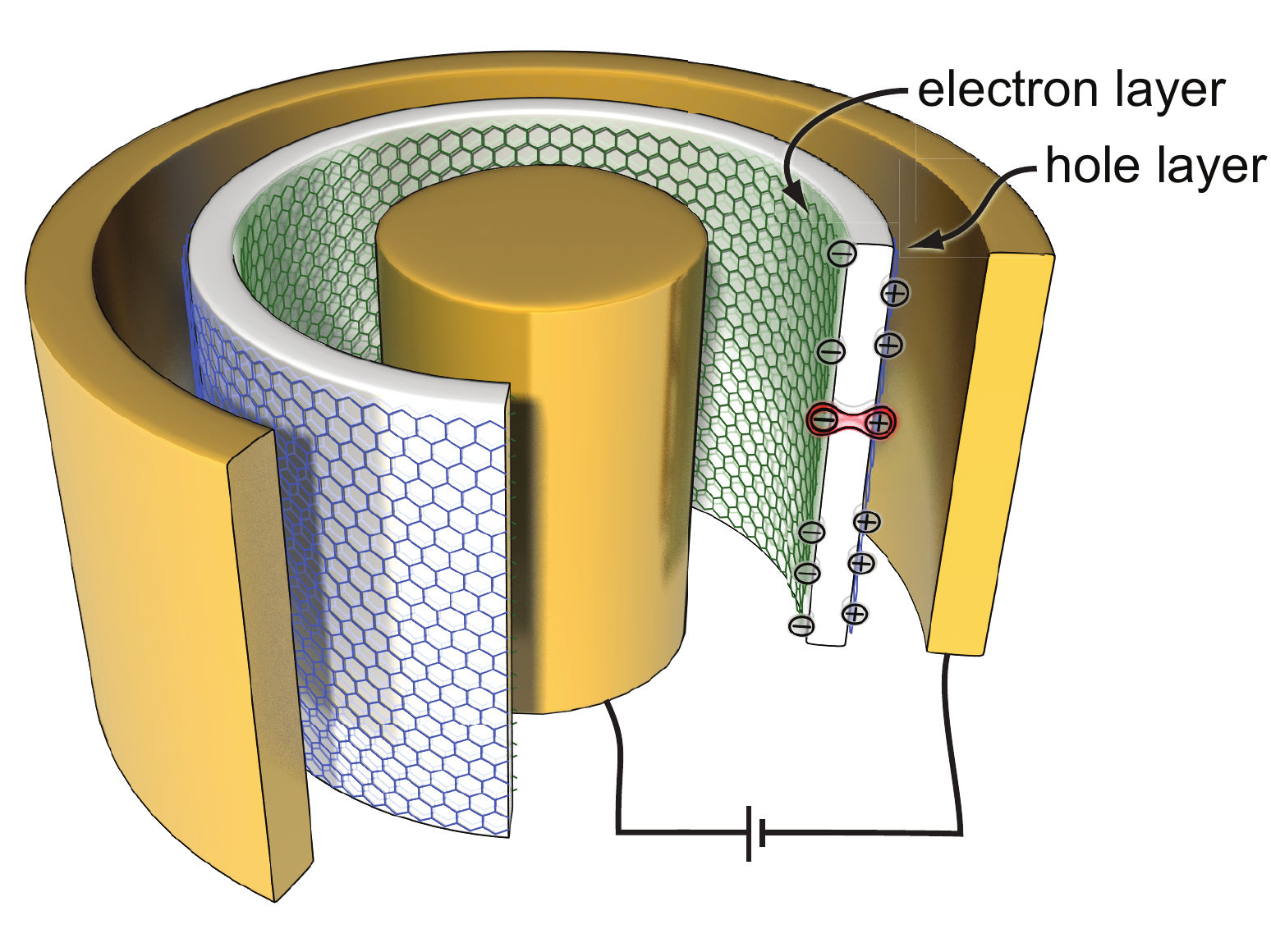}
 \caption{\label{fig3}Schematic representations of possible practical realizations of the concentric ring geometry comprised of p- and n-doped layers. Left: Using epitaxially grown complex oxide thin films. Right: Using doubly-gated graphene double layers.}
\end{figure}

\emph{Flux quantization effects.}\textemdash Let us now turn our attention to quantization effects. Imposing single-valuedness on the order parameter implies that for any given contour $C$ inside a superfluid $\oint_C \vec{\nabla} \phi \cdot \vec{dl} = \oint_C \vec{j} \cdot \vec{dl} = 2 \pi n$ where $n$ is an integer. Therefore, circular supercurrents must be quantized. In dipolar superfluids theories this would lead to two kinds of quantization effects. Dipoles pointing in the axial direction leads to the quantization of 'magnetic monopoles' \cite{Babaev:2008p4827, Seradjeh:2008p4898}, similar to electric line charge quantization in spin superfluids \cite{Leurs:2008p4904}. Here we are interested in devices where the electric dipoles point in the radial direction.

Consider a cylindrical device of radius $r$ consisting of two concentric layers as shown in Figure 2 with the electric dipole moment $\vec{d}$ of the excitons pointing in the radial direction. For this geometry the current-dependent term in the free energy can be written as
\begin{equation}
	F[\Psi] \sim \oint d\theta \left[ \frac{h}{e} \partial_\theta \phi
		- B_z 2 \pi r d \right]^2
	\label{FreeEa}
\end{equation}
where $\oint_C d\theta \partial_\theta \phi = 2 \pi n$ with $n$ integer valued and $B_z$ is the external magnetic field. Note that the flux going in between the two layers equals up to first order $\Phi = B_z 2 \pi r d$. Minimization of equation (\ref{FreeEa}) shows that current quanta can be induced by an axial magnetic field. In the absence of the external field, the current $\vec{j} \sim n$ induces a magnetic flux in between the layers, according to Amp\`{e}re's law (\ref{Ampere}), with a magnitude
\begin{equation}
	\Phi = \frac{h}{e} \, \chi_m \, n \equiv \Phi_0 \, \chi_m \, n.
\end{equation}
This is our central result: in the cylindrical double layer geometry, the magnetic flux going in between the sample layers must be quantized in units of $\chi_m$ times the fundamental flux quantum $\Phi_0 = \frac{h}{e}$. Notice that this flux quantization effect is quite different from the one realized in superconductors. In the double layer exciton condensate the supercurrent is induced by the magnetic field $\vec{B}$ rather than the gauge field $\vec{A}$ as in the London equation, while the quantized amount of flux equals $d \oint \vec{B} \cdot \vec{dl}$ instead of $\oint \vec{A} \cdot \vec{dl} = \int \int_\Sigma \vec{B} \cdot \vec{d\Sigma}$ for superconductors. In combination these two basic differences add up to an universal expression for the flux quantization $\Phi = \frac{h}{e^*}\, \chi_m \, n$ that applies to both superconductors and exciton condensates, where $e^* = -2e$ and $\chi_m = -1$ for superconductors. 

\emph{Phase slip and phase pinning.}\textemdash Is the strength of the condensate sufficient to trap the flux? When the external field is switched off the flux carrying state is metastable and the system can return to the ground state by locally destroying the condensate. The condensate can only be destroyed over lengths greater than the Ginzburg-Landau coherence length
\begin{equation}
	\xi = \frac{\hbar}{\sqrt{|2m^* \alpha |}}
\end{equation}
and consequently the energy required to break the condensate over a region $\xi$ wide along a cylinder of length $z$ is
\begin{equation}
	\delta F_b 	= \frac{1}{2} \hbar z \left( \frac{d\; \rho}{2m^* \mu} \right)^{1/2} B_c.
\end{equation}
Locally destroying the condensate is only favorable if this energy is lower than the energy stored in the magnetic field, which is $\delta F_m = \frac{B^2}{2 \mu} 2 \pi r d$. We conclude that a phase slip will not occur as long as the trapped magnetic flux $\Phi = \Phi_0 \chi_m n$ stays below a threshold value,
\begin{equation}
	\Phi^2 < \left( \frac{|\chi_m|}{2} \right)^{1/2} \Phi_0 B_c \; r d.
\end{equation}
where $B_c$ is the critical magnetic field. With the typical parameters stated above and $r = 100$ $\mu$m, the critical field must exceed $5$ nT to trap one flux quantum. Since the critical magnetic field of bilayer superfluids is proposed to lie in the orders of tens of Teslas\cite{Balatsky:2004p3613}, a phase slip is improbable.

Another possible complication is that annihilation of excitons by tunneling causes the phase to be pinned which introduces a threshold for the formation of stable currents. Microscopic tunneling can be incorporated via an extra term in the Ginzburg-Landau free energy
\begin{equation}
	F_t = - 2t \int d^2x \frac{|\Psi |}{L} \cos \phi,
\end{equation} 
where $L$ is the in-plane lattice constant and $t$ is a microscopic tunneling energy. This phase pinning lowers the energy of the state where no flux is trapped, which introduces a threshold for the trapping of magnetic flux quanta. It is only possible to trap $n$ magnetic flux quanta if the microscopic tunneling energy $t$ satisfies
\begin{equation}
	2t < n^2 \frac{\hbar^2}{2 m^* r^2} \; \sqrt{\rho} L.
	\label{TunnelingThreshold}
\end{equation}
This corresponds, given the typical parameters mentioned above, to $t < 0.3 $ peV (pico-electronvolt) for the first flux quantum.

In order to estimate a physical value for $t$, let us imagine that the device is fabricated from copper-oxide layers. The hopping energy in cuprates between two adjacent CuO$_2$ layers ranges from approximately $10^{-1}$ eV for LSCO compounds to $10^{-3}$ eV for Bi-based compounds \cite{GrayCooper,Clarke}. Let us now assume that the hopping energy between more distant CuO$_2$ layers falls off exponentially. A distance $d= 20$ nm between the hole and electron layer corresponds roughly to 30 CuO$_2$ layers, so that the tunneling energy equals $t \approx e^{-30} 10^{-3} = 10^{-16}$ eV. This estimate lies well below the maximum value of $t$ obtained in equation (\ref{TunnelingThreshold}).

\emph{Experimental realization.}\textemdash The experimental protocol to test the flux quantization is as follows: apply an axial magnetic field of magnitude $B_{\mathrm{ext}}$ above the critical temperature $T_c$, and cool the device below $T_c$ such that a circular current quantum is frozen in. The magnitude of the current is determined by the strength of the applied flux: if $\Phi_{\mathrm{ext}} < \frac{1}{2} \Phi_0$ no current is induced, for $\frac{1}{2} \Phi_0 < \Phi_{\mathrm{ext}} < \frac{3}{2} \Phi_0$ one current quantum is induced, etc. The magnetic field corresponding to $\frac{1}{2} \Phi_0$ is typically $B_{\mathrm{ext}} = 0.2$ mT. Upon removing the external magnetic field, a trapped flux equal to $\Phi_0 \chi_m n$ remains, corresponding to a field strength of $50$ nT. These numbers do not pose a problem of principle for the experimental realization of such a flux trapping device. 

Based on existing technology, one can envision various practical realizations of the concentric p-n doped ring geometry, while it is anticipated that further technology developments will create additional opportunities. Using p- and n-doped complex oxide compounds, such as cuprate perovskites, multilayer thin film structures can be fabricated in the desired ring geometry. Using the proven edge-junction technology \cite{Gao,Hilgenkamp} the structure sketched in Figure \ref{fig3} can readily be fabricated, by e.g., pulsed laser deposition and Ar-ion beam etching. As a barrier layer SrTiO$_3$ can be used, with a typical thickness of 10-100 nm, or another insulating oxide that grows epitaxially on top of the etched base electrode. To guarantee an epitaxial growth of all the layers, the angle $\alpha$ is best kept below about 25$^{\circ}$, but this does not fundamentally alter the physics of the flux quantization as presented in this letter.   

A second possible practical realization is based on double-side gated, double layer graphene. Recently, the growth of large area graphene films has been demonstrated on Cu foils, using a high temperature chemical vapor deposition process \cite{Li}. Interestingly, a continuous growth was achieved over grain boundaries and surface steps. From this it feasible to expect that one can also grow a closed graphene tube around a copper cylinder, which would basically be a carbon nanotube with predetermined radius. Covering this with an appropriate epitaxial barrier layer, e.g. 10 nm of Al$_2$O$_3$ and a second graphene sheet, which may also be grown by physical or chemical vapor deposition-techniques, would then result in the wanted concentric cylinder configuration. Subsequently, the copper can be etched away and the concentric cylinder can be transferred to an appropriate carrier, which can even be made out of plastic \cite{Bae}. This would straightforwardly allow realizing a doubly gated configuration as depicted in Figure \ref{fig3}.

\emph{Conclusion.}\textemdash We have shown that dipolar exciton condensates exhibit a novel form of magnetic flux quantization. Whereas the values for the magnetic flux quanta are reduced by a factor $2 \chi_m \approx 10^{-4} - 10^{-3}$ compared to the standard flux quanta in superconducting rings it is anticipated that the flux quantization is measurable using scanning SQUID microscopy. This would provide an unambiguous test for the macroscopic phase coherence associated with an exciton Bose-Einstein Condensate.

\emph{Acknowledgements.}\textemdash We thank A.~V. Balatsky, P.~H. Kes, J. van der Brink and P.~B. Littlewood for helpful discussions, and J. Huijben for designing the figures. This work has been supported by the Netherlands Organization for ScientiÞc Research (NWO).

\end{document}